\documentclass{raa}

\usepackage{graphicx,times}             %for PS/EPS graphics inclusion, new
\input{epsf.sty}                        %for PS/EPS graphics inclusion, old

\begin{document}

    \title{Flare processes evolution and polarization changes of fine structures
           of solar radio emission in the April 11, 2013 event
 %\,$^*$
 %\footnotetext{$*$ Supported by the National Natural Science Foundation of China.}
 }
 %   \subtitle{I. Place Your Subtitle Here}

    \volnopage{Vol.0 (200x) No.0, 000--000}      %%preserved for Editor. DOn't remove!
    \setcounter{page}{1}          %%starting page, preserved for Editor. DOn't remove!

    \author{Gennady Chernov\inst{1,2}, Robert Sych\inst{3,4}, Baolin Tan\inst{1}, Yihua Yan\inst{1}
     Chengming Tan\inst{1}, Qijun Fu\inst{1}, Marian Karlick\'y\inst{5} \and Valery Fomichev\inst{2}}

\institute{Key Laboratory of Solar Activity, National Astronomical Observatory, Chinese Academy
    of Sciences, Beijing 100012, China; {\it gchernov@izmiran.ru}\\
		\and Pushkov Institute of Terrestrial Magnetism, Ionosphere and Radio Wave Propagation, Russian
 Academy of Sciences (IZMIRAN), Troitsk, Moscow region, 142190, Russia\\
\and Yunnan Observatories, Chinese Academy of Sciences, Kunming 650011, China\\
\and Institute of Solar-Terrestrial Physics, Russian Academy of Sciences, Siberian Branch, Irkutsk 664033, Russia\\
 \and Astronomical Institute of the Academy of Sciences of the Czech Republic, Ondrejov 15165,
 Czech Republic}

    \date{Received~~2015 January 31; accepted~~2015~~August 31}

 \abstract{ The measurement of positions and sizes of radio sources in the observations of solar radio spectral fine structures in an M6.5 flare on April 11, 2013 were observed simultaneously by several radio instruments at four different observatories:
 Chinese Solar Broadband Radio Spectrometers at Huairou (SBRS/Huairou), Ond\v{r}ejov Radio
 spectrograph in the Czech Republic (ORSC/Ond\v{r}ejov), Badary Broadband Microwave
 spectropolarimeter (BMS/Irkutsk), and spectrograph/IZMIRAN (Moscow, Troitsk).
 The fine structures include microwave zebra patterns (ZP) fast pulsations, and fibers.
 They were observed during the flare brightening located at the tops of a loop arcade as
 shown in images taken by the extreme ultraviolet (EUV) telescope on board NASA's satellite
 Solar Dynamics Observatory (SDO) images. The flare occurred at 06:58 - 07:26 UT in solar
 active region NOAA 11719 located close to the solar disk center. ZP appeared near high
 frequency boundaries of the pulsations, and their spectra observed in Huairou and Ond\v{r}ejov
 agreed with each other in terms of details. At the beginning of the flare impulsive phase,
  a strong narrowband ZP burst occurred with a moderate left-handed circular polarization.
   Then a series of pulsations and ZPs were observed almost in an unpolarized emission.
   After 07:00 UT a ZP appeared with a moderate right-handed polarization. In the flare
   decay phase (at about of 07:25 UT), ZP and fiber bursts become strongly right-hand
   polarized. BMS/Irkutsk spectral observations indicated that the background emission
   showed left-handed circular polarization (similar to SBRS/Huairou spectra around 3 GHz
    in). However, the fine structure appeared in the right-handed polarization. The dynamics
  of the polarization was associated with the motion of the flare exciter, which was observed
  in EUV images at 171 \AA~ and 131 \AA~ by the SDO Atmospheric Imaging Assembly (AIA). Combining
  magnetograms observed by the SDO Helioseismic and magnetic Imager (SDO/HMI) with the homologous
  assumption of EUV flare brightenings and ZP bursts, we deduced that the observed ZPs
  correspond to the ordinary radio emission mode. However, future analysis needs to verify
  the assumption that zebra radio sources are really related to a closed magnetic loop, and
  are located at lower heights in the solar atmosphere than the source of pulsations.
  }

    \authorrunning{Gennady Chernov\inst{1,2}, Robert Sych\inst{4}, Baolin Tan\inst{1} et al.} %author_head in
    \titlerunning{Flare processes evolution and polarization changes}  % titl

    \maketitle
 %% The author head (on even pages) and the title head (on odd pages) will be
 %% automatically extracted from \author{} and \title{}. Whenever the title is too long,
 %% you will be asked to supply a shorter one by inserting either \authorrunning{} or
 %% \titlerunning{} before \maketitle. Anyway, you can specify your own heads.
 %%
 %%
 %% Note: In the following text body of your manuscript, please note several differences from
 %%       other major journals:
 %% (1) \subsection{Please Capitalize the First Letter of Each Notional Word in Subsection Title}
 %% (2) Please Capitalize the First Letter of Each Notional Word in all tables' captions               es

 %
 %________________________________________________ sections below
 %
 \section{Introduction}           %% first-level sections will be auto-capitalized
 \label{sect:intro}

   Zebra structure (zebra or zebra pattern - ZP) is the most intriguing fine structure in
   the dynamic spectra of solar and stellar radio bursts. It consists of several almost
   parallel stripes in emission and absorption against the background of the solar radio
   broadband type IV continuum emission. In particular, the microwave ZP supplies original
   information about the flaring source region where the primary energy is released, such as
   the magnetic fields, particle acceleration and plasma instabilities. The nature of ZP
   structures has remained a subject of wide discussion for more than 40 years. More than
   ten radio emission mechanisms were proposed for its explanation.  The statistics and
   classification of the microwave ZPs were presented recently (Tan et al. 2014a). The
   history of observations and of theoretical models were assembled in reviews by Chernov
   (2006), Zlotnik (2009) and Chernov (2011).
        Most often in the literature, the mechanism based on a double plasma resonance
        (DPR) is discussed  (Kuijpers, 1975; Zheleznykov and Zlotnik, 1975a,b; Kuijpers
        (1980); Mollwo, 1983; 1988; Winglee and Dulk, 1986). It assumes that the upper
        hybrid frequency ($\omega_{UH}$)
  in the solar corona becomes resonant at a multiple of the electron-cyclotron frequency:
     \begin{equation}
     \omega_{UH} = (\omega_{Pe}^2 + \omega_{Be}^2)^{1/2} = s\omega_{Be}
     \end{equation}
     where $\omega_{Pe}$-electron plasma frequency, $\omega_{Be}$- electron cyclotron
     frequency and s- integer harmonic number.
     This mechanism experiences a number of problems with the explanation of the dynamics
     of zebra stripes.

     Chernov (1976; 1990) proposed an alternative mechanism for ZP: the coalescence
     of plasma waves ($\it l$) with whistlers $\it(w)$, $\it l + w \to t $
    (Kuijpers, 1975). In this unified
     model the formation of ZP was attributed to the oblique propagation of whistlers,
     while the formation of stripes with a stable negative frequency drift (the fiber
     bursts) was explained by the ducted propagation of whistlers along a magnetic
     trap. This model explains not only some thin effects (sharp changes in frequency
     drift, a large number of stripes, frequency splitting of stripes, and superfine
     millisecond structure) but the occasionally observed transformation of the ZP
     stripes into fibers and vice versa, and also the synchronous variations in the
     frequency drift of stripes with the spatial drift of the radio sources. However,
     a certain boom of new models has been proposed (see more details Chernov (2006; 2014).
      The most crucial parameter for selecting the theoretical mechanism is
      the polarization of radio radiation in combination with radio source
      structure. The polarization mode is only defined when the probable
      radio source position can be identified on magnetograms. In most cases
      the ordinary wave mode was found for ZP, fiber bursts and fast radio
      pulsations. The model of ZP in conditions of the double plasma resonance
      (DPR) (Zlotnik, 2009) proposes the ordinary wave mode, and radio source
      must be stationary. In the model with whistlers (Chernov, 2006) the radio
      emission of ZP and fiber bursts is also related to the ordinary mode but
      the radio source should be moving. However in an event on 23 January 2003,
      Altyntsev et al. (2005) found an extraordinary mode using positional
      observations, therefore they related ZP with the mechanism of Bernstein modes.

So far, solar cycle 24 has shown a lack of major flares. However, small flares
presented a number of enigmas in radio emission with fine structures.  Among
several recent events, the most interesting is the M 6.5 flare that occurred at
06:58 - 07:26 UT on April 11, 2013.

In this event, we observed a very rare case of changes in the polarization of
the radio fine structure together with simultaneous observations of the flare
dynamic process using SDO/AIA at several EUV lines (Pesnell et al. 2012; Leven
et al. 2012), in hard X- ray images from the Ramaty High Energy Solar Spectroscopic
Imager (RHESSI) (Hurford et al. 2002; Krucker, Lin, 2002) and SDO/HMI magnetograms
(Schou et al. 2012). During this event, the polarization was different in each
appearance of the fine structure and even had different signs.

Thus, this event offers a very exceptional possibility to determine the radio
emission wave mode of ZPs. We will use the classical definition of the radio
emission mode in radio physics and radio  astronomy: the rotation of the electric
vector of the extraordinary wave mode coincides with the direction of rotation of
electrons in the magnetic field on the right spiral (clockwise rotation when viewed
along the positive field).  Thus, right handed circular polarization above positive
(North) magnetic polarity corresponds to the extraordinary wave mode, and is left
handed with respect to the ordinary one. So, we could use a simple abbreviation for
the emission modes: RoS, LeS, ReN and LoN, where R and L are Right and Left signs of
the polarization respectively, S and N are magnetic polarities, and o and e are
ordinary and extraordinary wave modes respectively.

 %% Authors can give a citation as 'Michel et al. 1992'.
 %% You may also use \cite, \citep and \citet for citation, and use Table~1 or Figure~1
 %% and so forth. Using \ref and \label for cross-references of Tables/Figures
 %% is a good way in adjusting/adding/removing text, tables or figures.

 \section{Observations}
 \label{sect:Obs}

 \subsection{General data }

    For the first time, there are several radio telescopes at four different observatories
    that observed the radio bursts of a flare event simultaneously on 2013 April 11
    (07:00 -- 07:26 UT). These telescopes include: Chinese Solar Broadband Radio
    Spectrometers at Huairou (SBRS/Huairou), Ond\v{r}ejov Radio Spectrographs in the
    Czech Republic (ORSC/Ond\v{r}ejov), Badary Broadband Microwave Spectropolarimeter
    (BMS/Irkutsk), and spectrograph/IZMIRAN. SBRS/Huairou, spectrographs/Ondrejov,
    spectrograph/Badary, and spectrograph/IZMIRAN.

We used broadband spectrograms of: the SBRS/Huairou in the range 2.60 - 3.80 GHz
(with an antenna diameter of 3.2 m, cadence of 8 ms, frequency resolution of 10 MHz,
Fu et al (1995; 2004)); high frequency part of ORSC/Ond\v{r}ejov (Jiricka et al, 1993)
at frequencies of 2.00 - 5.00 GHz with a frequency resolution of 12 MHz and time
resolution of 10 ms in total radio flux; the  broadband microwave spectropolarimeters
Badary in the range 4.0-7.0 GHz with  resolutions 100 MHZ and 20 ms (Zhdanov and Zandanov
2011); spectrograph/IZMIRAN in the meter range 25 - 270 MHz (Gorgutsa et al. 2001).

 The whole flare process on April 11, 2013 took place in solar active region NOAA
 11719 which was located very close to the center of the solar disk (N09E12). It was
 a two-ribbon M6.5 flare that started at 06:58 UT, reached its maximum at 07:16 UT,
 and ended at 07:26 UT.

   \begin{figure}
    \centering
   \includegraphics[width=12cm, angle=0]{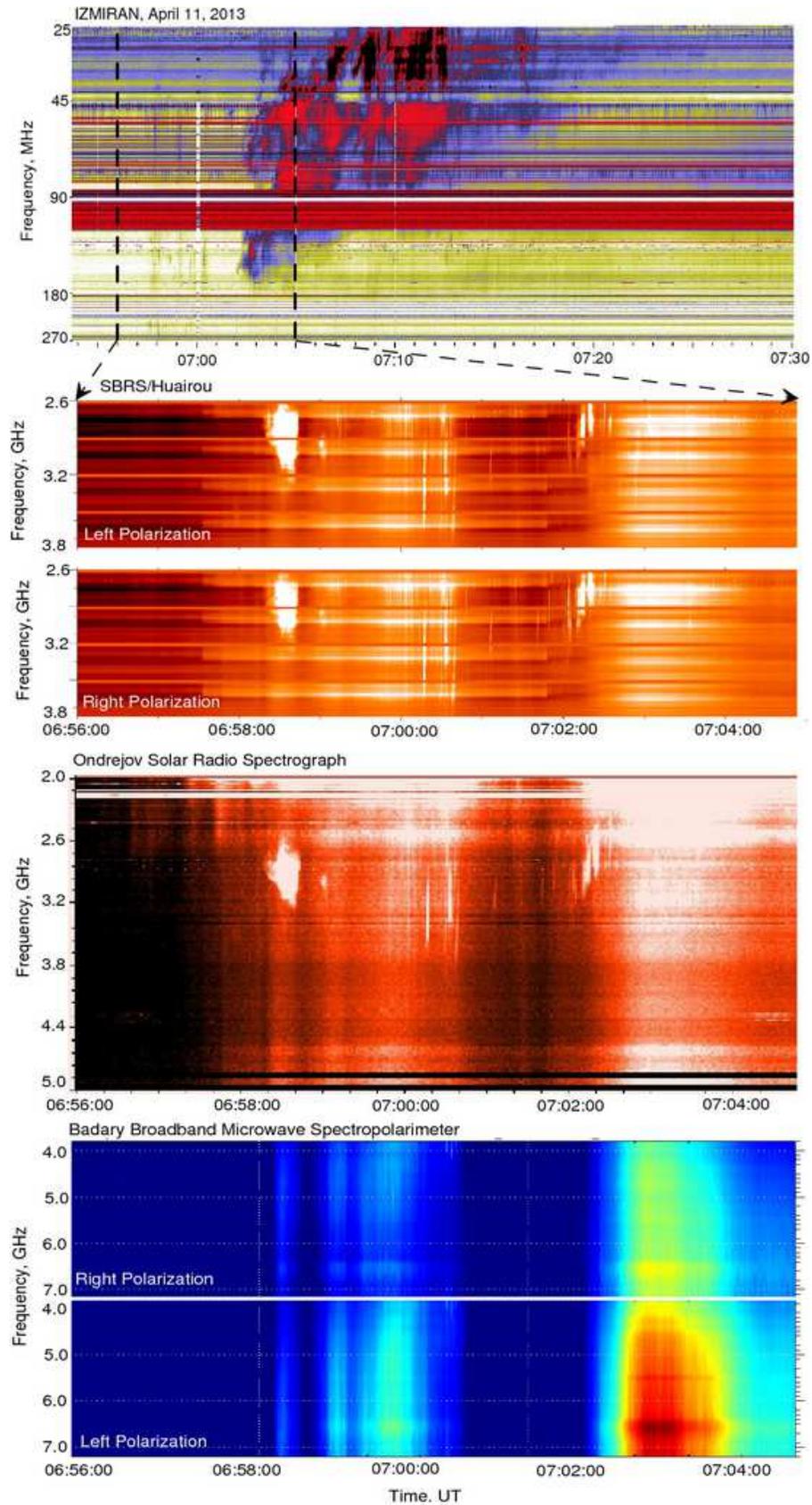}
    \caption{General view of radio spectra from the event on April 11,
    2013 observed at four observatories.}
    %\label{Fig:demo1}  Demo1: Fig. 1
    \end{figure}

 Tan et al. (2014b) reported the first ZP burst which appeared in a strong narrow band
 burst around 3 GHz at the beginning of the flare (06:58:30 UT) (see middle panels of
 Figure 1 observed by the SBRS/Huairou and ORSC/Ondrejov). Furthermore, ZP was observed
 to accompany to all consecutive pulsations. According to the observation data obtained
 by BMS/Irkutsk in the range 4 - 7 GHz, the continuum radio emission was strongly
 left-handed polarized (bottom spectrum in Figure 1). The radio burst began almost
 simultaneously in meter and centimeter ranges: a strong type II burst began at 07:02
 UT after a group of type III bursts at 06:57 UT (top IZMIRAN's spectrum in Figure 1).
 There was no strong continuum emission with any fine structure in the meter range.
 It is possible that the magnetic force lines were open (not a magnetic trap) after
 the observed ejection of CME. According to the catalog data of SOHO/LASCO C2, the
 extrapolated straight line in the diagram of height- time for this CME (of halo type)
 notes the beginning of a disturbance at 06:50 UT. Approximately at the same moment it
 is possible to trace the beginning of ejection in the EUV coronal lines 171 \AA~ and 195 \AA.
 Therefore a corresponding shock wave could be driven by the CME front. The beginning of
 the event and related instrumentations were described by Tan et al. (2014).

 The two top panels in Figure 2 show two images in hard X- ray observed by RHESSI.
In the preflare phase, the hard X- ray source had a loop configuration composed of three
parts: two footpoint sources and one looptop source. However, near the flare maximum only
the looptop source remains, like a loop arcade. Such a change in the source structure is
very typical for the magnetic reconnection. The brightest flare kernels indicate the
locations of maximal acceleration of fast particles.

    \begin{figure}
    \centering
   \includegraphics[width=\textwidth, angle=0]{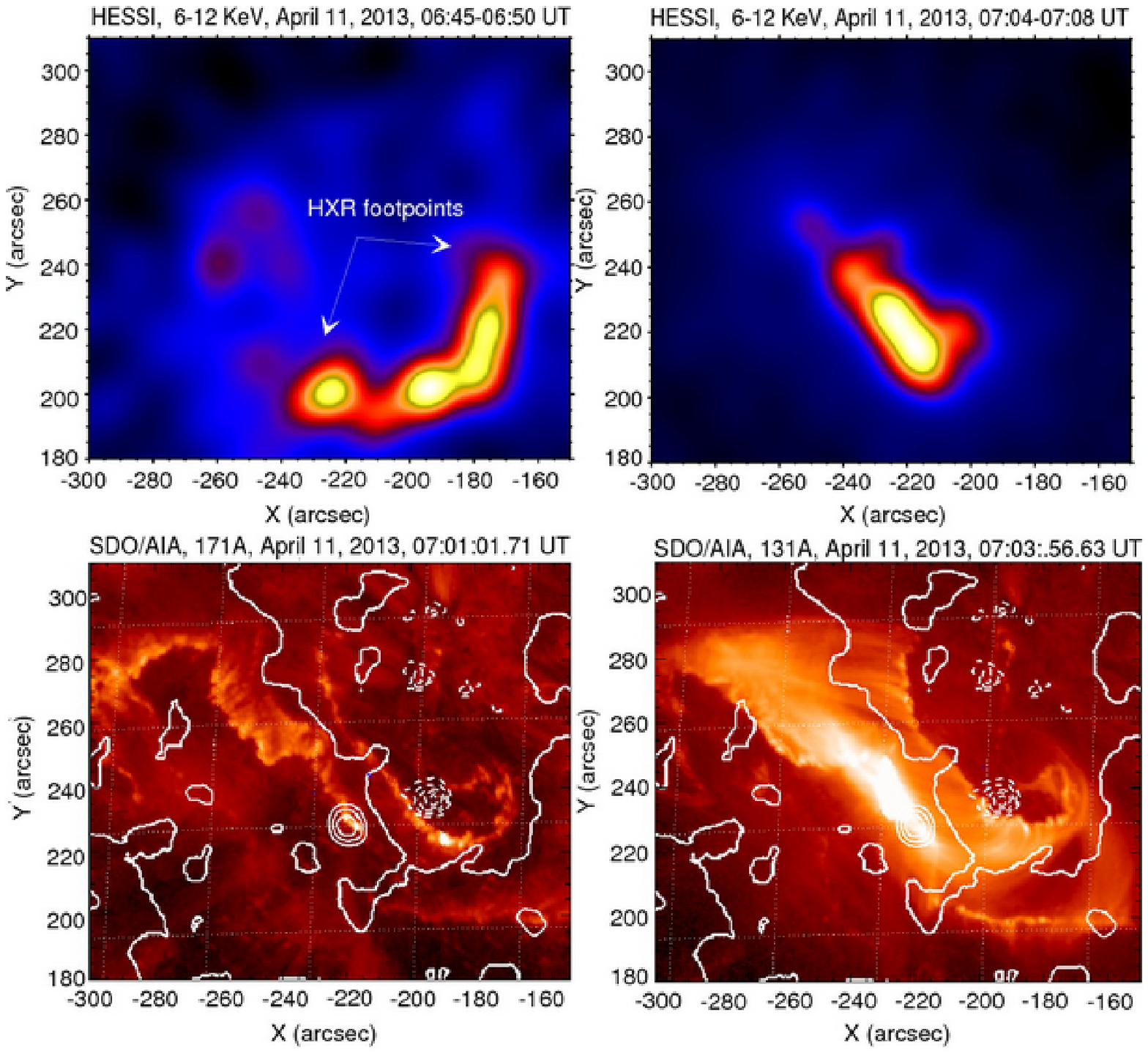}
    \caption{
 RHESSI hard X- ray images (top): In the preflare phase the hard X- ray
 source had a loop configuration, and then, on the growing phase of the
 event, only a looptop source remains at the arcade. Bottom panel: two
 frames from the movie from SDO/AIA 171 \AA~ (left) and 131 \AA~ (right)
 superimposed on a background of the SDO/HMI magnetogram: two flare
 ribbons are located on both sides of the magnetic neutral line (white
 thick solid line).}
    \end{figure}

Two bottom panels in Figure 2 show the position of the two flare ribbons
visible in two lines of SDO/AIA 171 \AA~ (coronal temperature $10^{6}$ K) and 131 \AA~
(hot plasma $10^{7}$ K) against the background of the SDO/ HMI magnetogram in the
 impulsive phase of the flare (the leading spot with South magnetic polarity
 is shown by the dotted line). The cadence of HMI data is of 45 s. The main
 feature of this flare is that flare brightenings arose alternately in both
 flare ribbons located on opposite sides of the neutral magnetic line (white
 thick solid line in Figure 2). The right frame in the hot line 131 \AA~ shows
 that around 07:03:56 UT the maximum energy release taken place in the left
 flare ribbon in North magnetic polarity.

  \begin{figure}
  \centering
  \includegraphics[width=\textwidth, angle=0]{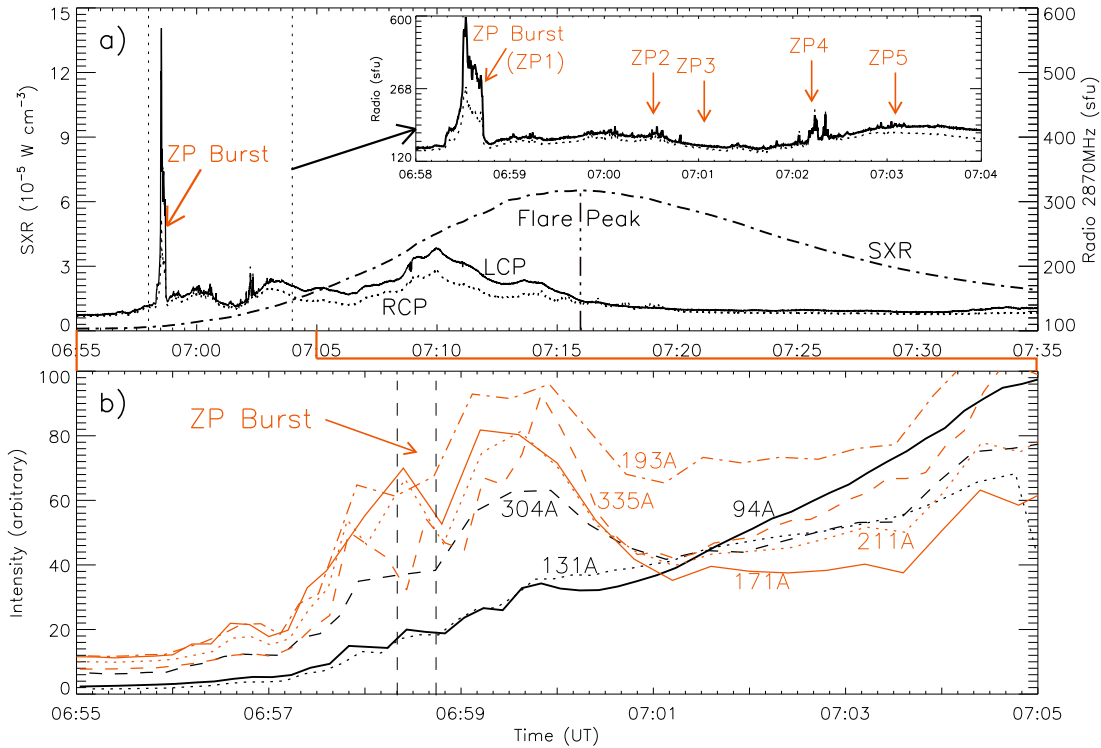}
  \caption{a: time profiles of radio emission intensities at 2.87 GHz at
  left-and right-handed circular polarization (LCP and RCP) observed by
  SBRS/Huairou, and the GOES soft X-ray 1-8 A. The increased fragment of
  profiles with the ZP at the beginning of the event is isolated at the top.
  The moments ZP1- ZP5 will be presented in Fig. 4 - 8.
   b: profiles of the integral EUV brightness; bottom panel shows
  an obvious enhancement at wavelengths of 171 \AA, 211 \AA, and 335 \AA~ (coronal
  temperature around $10^{6}$ K) at the moment the ZP appeared.}
  \end{figure}

Radio intensity profiles and the profiles at several EUV lines are shown in Figure 3.
According to GOES data the maximum of the event happened in soft X- ray around 07:17
UT but in radio emission at 2.87 GHz was around 07:10 UT.
 The first strong ZP burst occurred just at the beginning of the flare (Figure 4).
 Tan et al. (2014) described it and correlated it with a sudden EUV flash observed
 by SDO/AIA images. However, in several subsequent radio maxima (see profiles in
 Figure 3) many complex fine structure elements were observed with complex behavior
 in the polarization, although with less intensity.

 The polarization value was very variable and even had different signs. As show
 RCP and LCP channels of SBRS/Huairou in middle panels in Figure 1, the
 polarization of the strong burst at the beginning, after 06:58:30 UT was
 moderate left-handed, but the subsequent pulsations after 07:00 and 07:02 UT
 had moderate right-handed polarization. In the absence of positional radio data,
 we use SDO/AIA EUV images at a cadence of 12 s and pixel size of $0.6^{\prime\prime}$ at seven
 EUV wavelengths, and sometimes they coincided very well with moments of the fine
 structure. In such cases we propose that the radio source in microwaves could be
 located above these new flare kernels in EUV images. We could also compare these
 locations with SDO/HMI magnetograms (Figure 2) to determine the radio wave mode.

 \subsection{Radio spectral fine structures and dynamics of flare processes}

      The first ZP (Figure 4) occurs just at the beginning of the flare in the strong
      narrowband burst around 2.9 GHz shown in the spectrum of ORSC/Ond\v{r}ejov in
      Figure 1. Its maximum emission flux at 06:58:30 UT was very strong, it exceeded
      the background flaring continuum intensity by several times (Figure 3 and 4).
      Tan et al. (2014) showed that this ZP was simultaneously observed with ORSC/Ondrejov
      spectrograph and all details (stripe separations, brightness, duration, the arched
      shapes of each stripe and the spiky super structure) are strictly identical to the
      spectrogram obtained by SBRS/Huairou. The middle panel in Figure 4 shows moderate
      left-handed polarization, of $35 \pm 5\%$. The bottom-left panels in Figure 4 show a strong
      new flare kennel in the left flare ribbon (marked by an arrow) in the image of the
      Fe IX line at 171 \AA~ with a coronal temperature around $10^{6}$ K. From the images of a
      hot line at 131 \AA~ (right panel in Figure 4) we can also see the appearance of hot
      plasma (around $10^{7}$ K) in the same place (marked by an arrow). Profiles in Figure
      3b only show rising intensity of this line. Moderate Left polarization can be
      explained by a bright flare core in the left flare ribbon in the North magnetic
      polarity (ordinary wave mode).

    \begin{figure}
    \centering
    \includegraphics[width=14cm, angle=0]{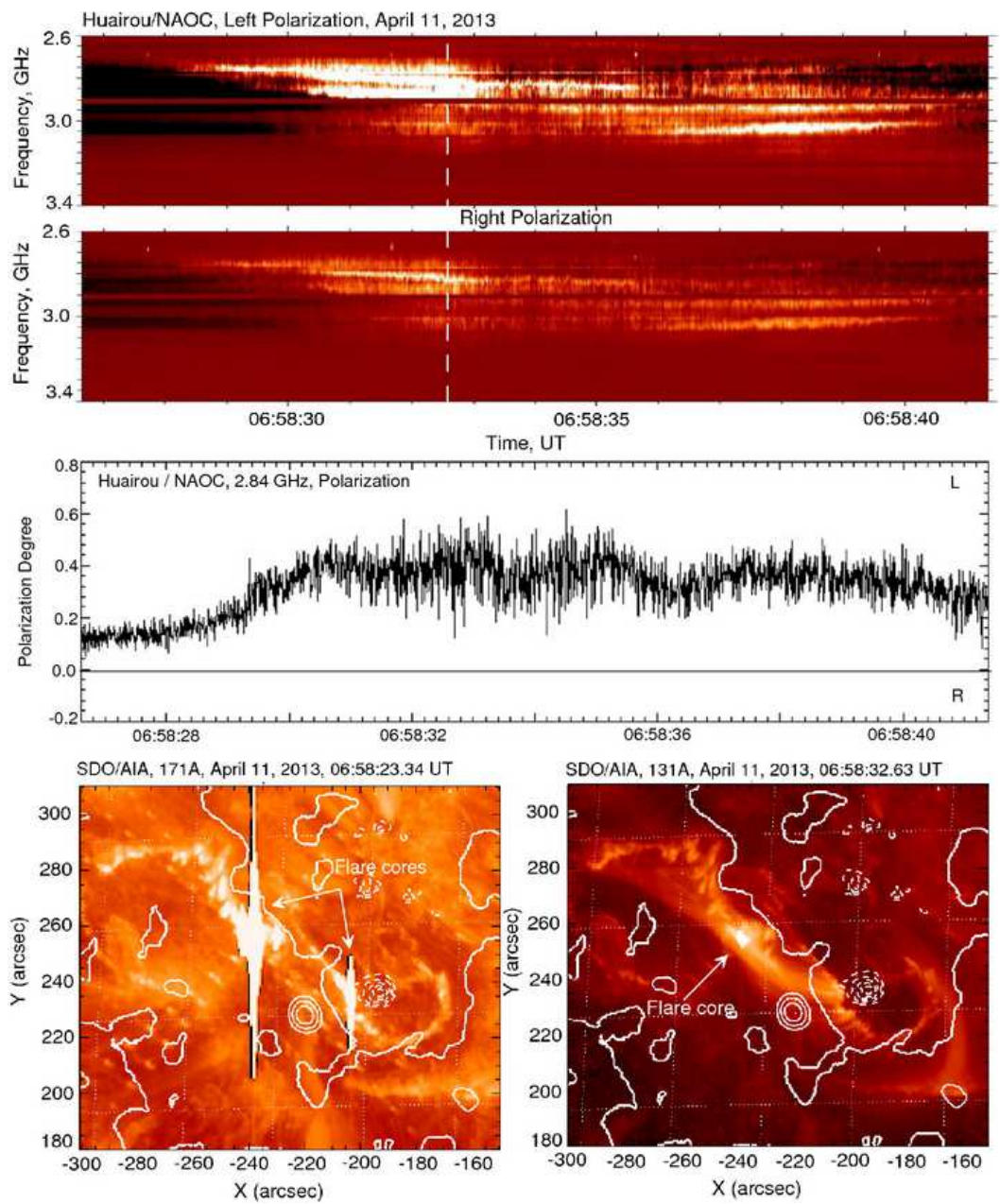}
    \caption{Strong ZP (the moment ZP1 in Fig. 3) at the beginning of the event that
    occurred on 11 April 2013 as registered by the SBRS/Huairou, discussed in Tan et al.
    (2014). Middle panel: polarization profile at 2.84 GHz. In the bottom: two frames
    from the movie of SDO/AIA 171 \AA~ (left) and 131 \AA~ (right). The dotted vertical
    line in the spectrum shows the moment displayed in the right frame at 131\AA.
    Moderate left polarization can be explained by a hard flare core in the left flare
    core in the left flare ribbon with N magnetic polarity (ordinary wave mode)}.
   \end{figure}

Two minutes after this first ZP, a series of pulsations with ZP of five second duration were
observed (Figure 5) but almost five times less intensity (see profiles in Figure 3a). However,
 the main distinction from the first ZP is changing of the polarization sign at the time of
 07:00:32.2 UT (from left- to right-handed) and further small oscillations about zero level
 of $\pm 7-10\%$. The level of a continuum is subtracted with the polarization
 calculation, therefore in the absence of a strong signal the error of polarization
 calculation increases. Such errors are removed by a digital filter. We can see enough narrow
 band pulsations with a periodicity of about 0.5 s, which are accompanied by several
 stripes in the form of ZP at their high frequency edge. ZP stripes (two or five) are
 absent between pulsations, therefore a relation exists between sources of pulsations and ZP

     The bottom left panel of Figure 5 in the passband of 171 \AA~ shows that the flare
     kernels in the left flare ribbon decreased but increased in the right ribbon (marked
     by an arrow), and in the hot plasma image at 131 \AA~ (right panel) a new flare feature
     appeared between the flare ribbons (marked by an arrow). It is evident from the movie
     frames that at the moment the main flare source appeared, a  loop arcade was located at
     the top between the flare ribbons. Thus, the source was found to be shifted to the
     leading spot of South magnetic polarity and this explains the weak polarization that
     even have an opposite Right- handed sign (which corresponds to the ordinary wave mode).

     About 40 s after the pulsations with ZP, new clear stripes of ZP appeared (Figure 6).
     The polarization profile shows very weak Right-handed polarization of 10\%. Although
     several new bright points appeared in both flare ribbon at 171 \AA~ image (i.e. in
     the footpoints of the loop arcade shown by arrows), the bright tops of the arcade
     in the hot 131 \AA~ image became broader, which shows the place with maximal energy
     release. In that case it does not make sense to estimate the wave mode. The frequency
     separation between ZP stripes grows with frequency from 60 to 90 MHz.

    \begin{figure}
    \centering
   \includegraphics[width=\textwidth, angle=0]{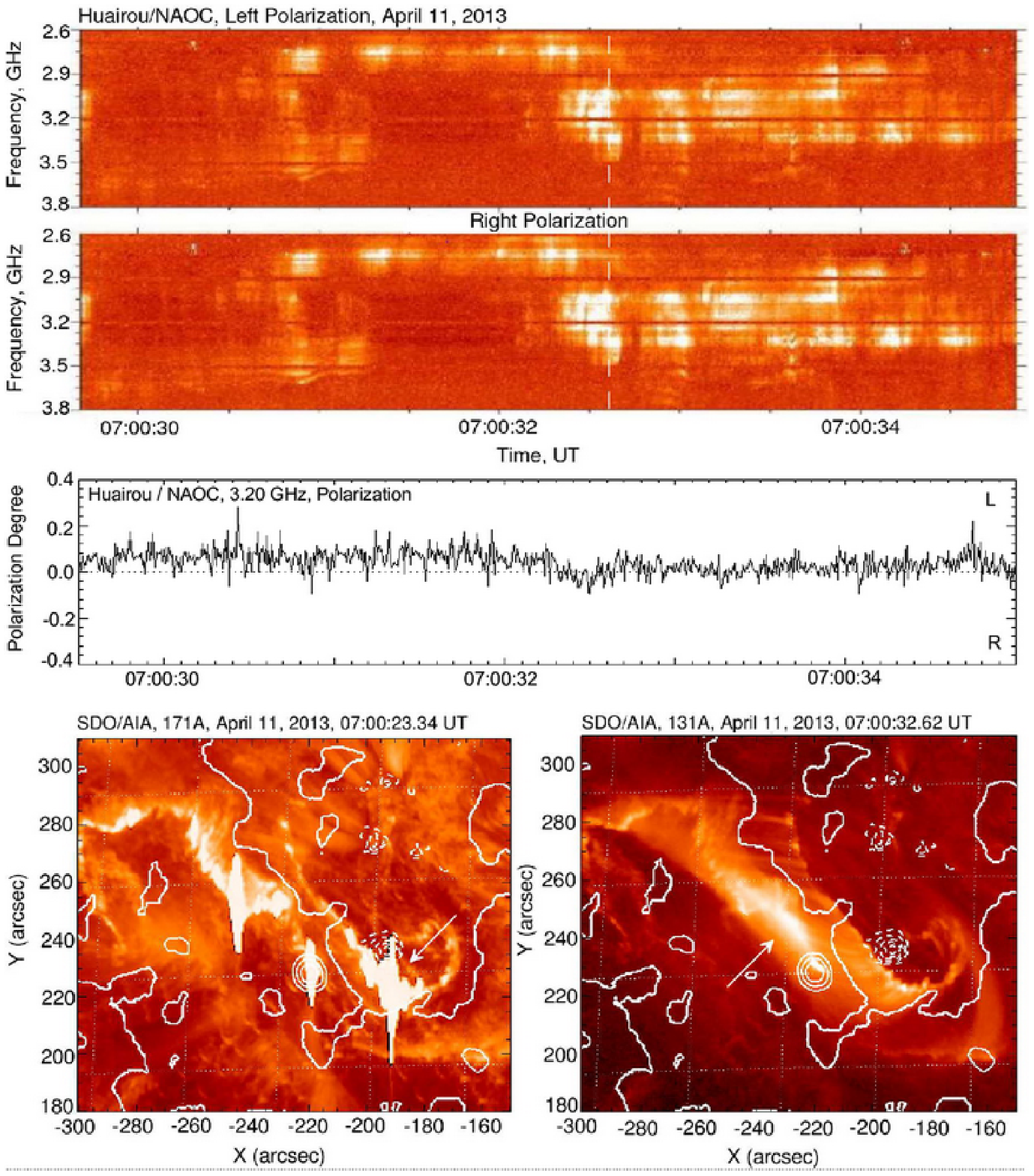}
    \caption{ Zebra pattern at the high frequency edge of the pulsations as registered by
    the SBRS/Huairou at 07:00:30 UT (the moment ZP2 in Fig. 3). The polarization profile
    in the middle panel shows change of the polarization sign at the time of 07:00:32.2 UT
    (from left- to right-handed) and further small oscillations about zero level of $\pm 7-10\%$.
    In the bottom: two frames from the movie from SDO/AIA 171 \AA~ (left) and 131 \AA~ (right).
    The dotted vertical line in the spectrum shows the moment displayed in the right frame in
    131 \AA~ line.}
    \end{figure}

One minute later, at 07:02:10 UT the most significant parts of spectra from this event appear,
and once more there are pulsations with ZP in their high frequency edge (Figure 7). The emission
is limited to high frequencies by several zebra stripes. As the left panel in 171 \AA~ shows,
in the left flare ribbon the flare kernels decreased and we see two bright and approximately
identical features in the footpoints of the arcade shown by arrows.  The movie in the
chromospheric images at 1600 \AA~ shows that a new disturbance began in the right flare ribbon.
According to the right frame in 131 \AA~ with hot plasma at the top of the flare arcade, the
flare expanded in the south-west direction (marked by an arrow) along both the flare ribbon
above the south magnetic polarity. Weak Right polarization above the south magnetic polarity
corresponds to the ordinary wave. The radio source of ZP is possibly related to a closed
magnetic loop, lower than the source of pulsations as well as for the first similar structure
that appears in Figure 5.

  \begin{figure}
  \centering
  \includegraphics[width=\textwidth, angle=0]{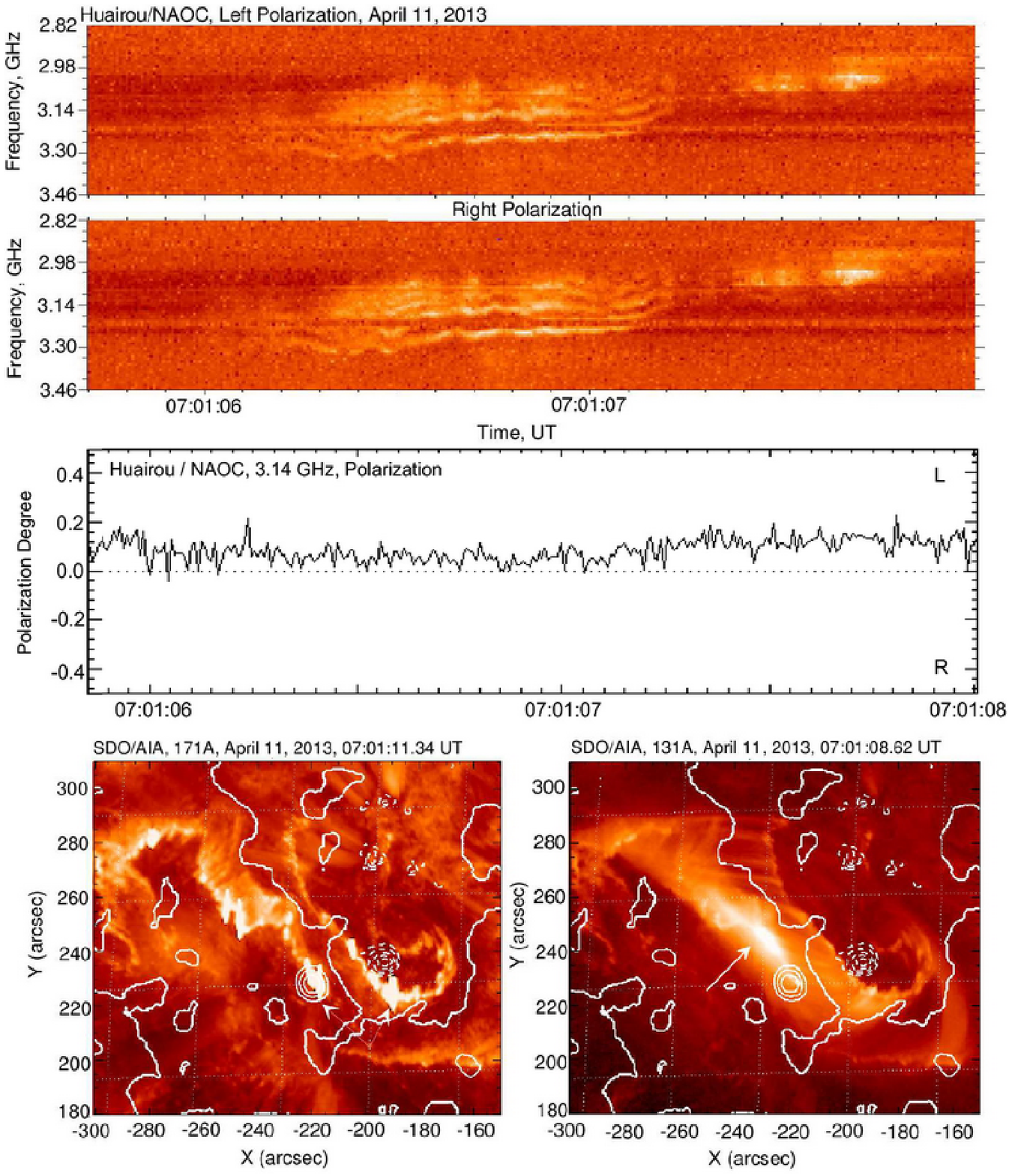}
  \caption{ Zebra pattern with weak Left polarization registered by the SBRS/Huairou at 07:01:06
  UT (the moment ZP3 in Fig. 3). The polarization profile in the middle panel shows weak
  Right-handed polarization of $10 \pm 5\%$. In the bottom: two frames from the movie of
  SDO/AIA  at 171 \AA~ (left) and 131 \AA~ (right).}
  \end{figure}

     Unlike Figure 5, these pulsations have, in addition to a period of about 0.5 s, an
     additional short period of about 0.15-0.20 s. The bandwidth of pulsations of about
     400 MHz remains more or less the same, but pulsations slightly drift to lower
     frequencies. All these peculiarities coincide with the fine structures in the Ond\v{r}ejov
     spectrum. Figure 7 also shows that the emission is absent at frequencies higher than
     3.3 GHz. However, a reduction in the continuum emission at lower frequencies began more
     than one minute before, at 07:01 UT (see Figure 1).
     One minute after the pulsation, at 07:03:03 UT a new ZP appeared with strong Left-handed
     polarization (Figure 8). New flare brightening occurred at 171 \AA~ in the left flare
     ribbon (marked by an arrow)  above the North magnetic polarity, and the bright arcade
     in 131 \AA~ also shifted to the left ribbon (marked by an arrow). Moderate Left-handed
     polarization (about of 25\%) above the North magnetic polarity corresponds to the
     ordinary wave mode.

     In the decay phase (after 07:16 UT to approximately 07:25 UT), the zebra and fiber
     bursts appeared with strong right handed polarization (see bottom profiles in Figure
     3a). The flare area expanded to the right flare ribbon (to the South magnetic polarity).
     The ordinary radio emission mode remains. The movie in 131 \AA~ after 07:00 UT with
     enhanced magnetic loops shows a restructuring of loops throughout the flare ribbons.
     The reconnection continues in the middle part of the flare ribbons, and the hot plasma
     ($10^{7} K$) covered the largest part of the flare area around 07:04:20 UT (the maximum
   of the intensity profile for the 131 \AA~ line in Figure 3b). During this interval,
   the movie also shows a rain of matter falling downwards, possibly after the first
   ejection around 06:50 UT.

     According to spectral data from Badary, the continuum emission has strong Left handed
     polarization (see Figure 1). However, sometimes we see the small bursts with Right
     handed polarization as is shown in Figure 9. Double bursts have a moderate Right-handed
     polarization (middle panel), and the polarization of the background continuum is at a
     noisy level because the continuum emission is subtracted with the calculation of
     polarization. In frame 171 \AA~ (coinciding with the time of the radio bursts) the
     two bright points appeared in the right flare ribbon. The frame of the chromospheric
     line at 1600 \AA~ shows that a fast disturbance was really initiated in the same place
     above the South magnetic polarity, and the radio emission can be related to the
     ordinary wave mode.  Large flare areas in the left ribbons shown by arrows in the
     bottom panels were more long-lived and they could be responsible for the continuum
     emission that had Left-handed polarization.

   Thus fast changes in the polarization were related to the dynamic motion of the flare
   cores between flare ribbons. In all possible cases (when the polarization was remarkable),
    we found that the polarization of the observed radio fine structures corresponds to the
    ordinary radio emission mode.

 \begin{figure}
 \centering
 \includegraphics[width=\textwidth, angle=0]{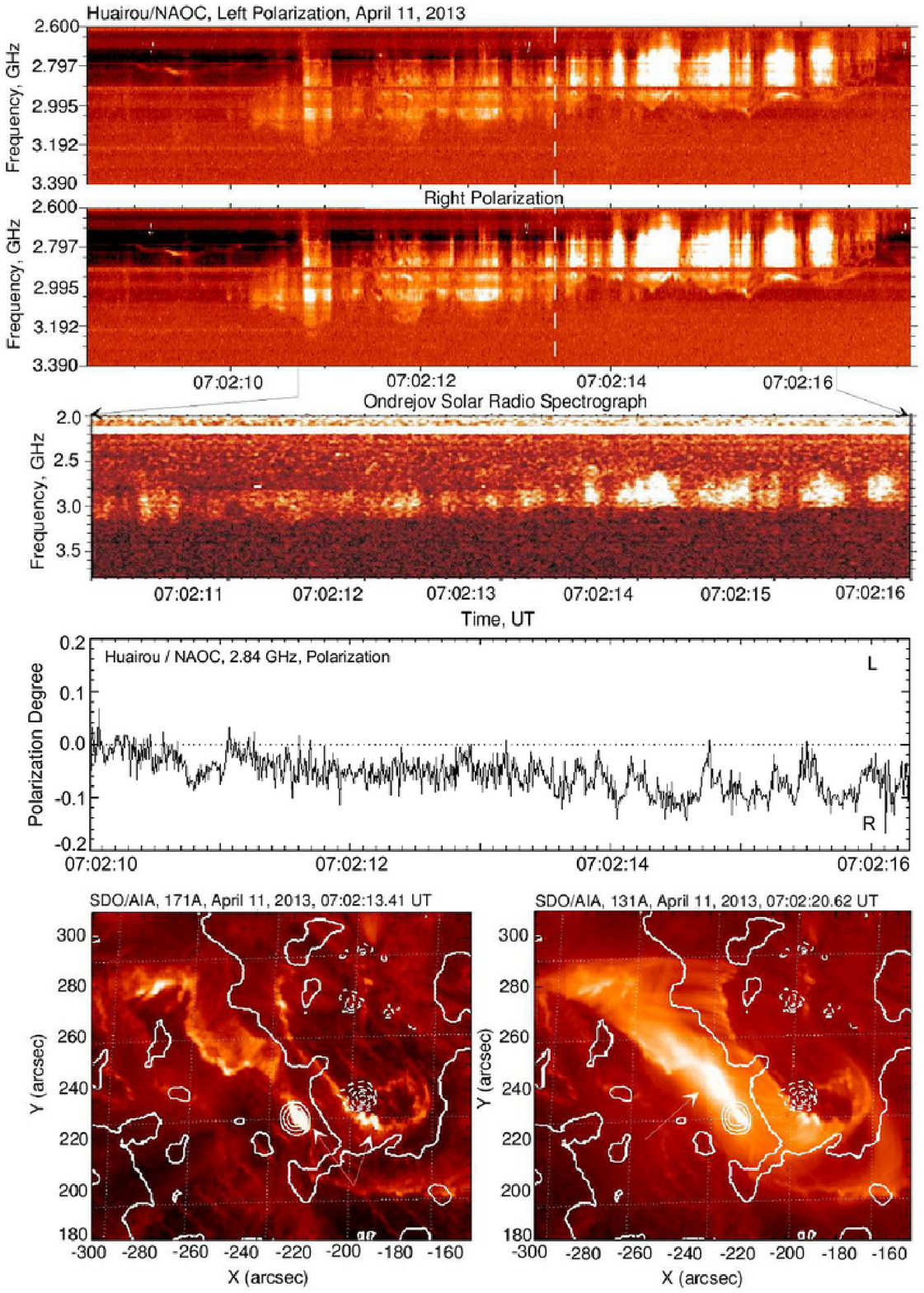}
 \caption{Fast pulsations during 8 sec registered by the SBRS/Huairou in the 11 April 2013
 event. Spectra of pulsations and ZP (the moment ZP4 in Fig. 3) coincide in observations
 from the Huairou and Ond\v{r}ejov observatories. The polarization profile at 2.84 GHz of
 SBRS/Huairou shows weak right-handed polarization of $8\pm4 \%$. Bottom: two frames from
 the movie of SDO/AIA at 171 \AA~ (left) and 131 \AA~ (right). The dotted vertical line in
 the spectrum shows the moment displayed in the left frame in 171 \AA~ line.}
 \end{figure}

  \begin{figure}
 \centering
 \includegraphics[width=\textwidth, angle=0]{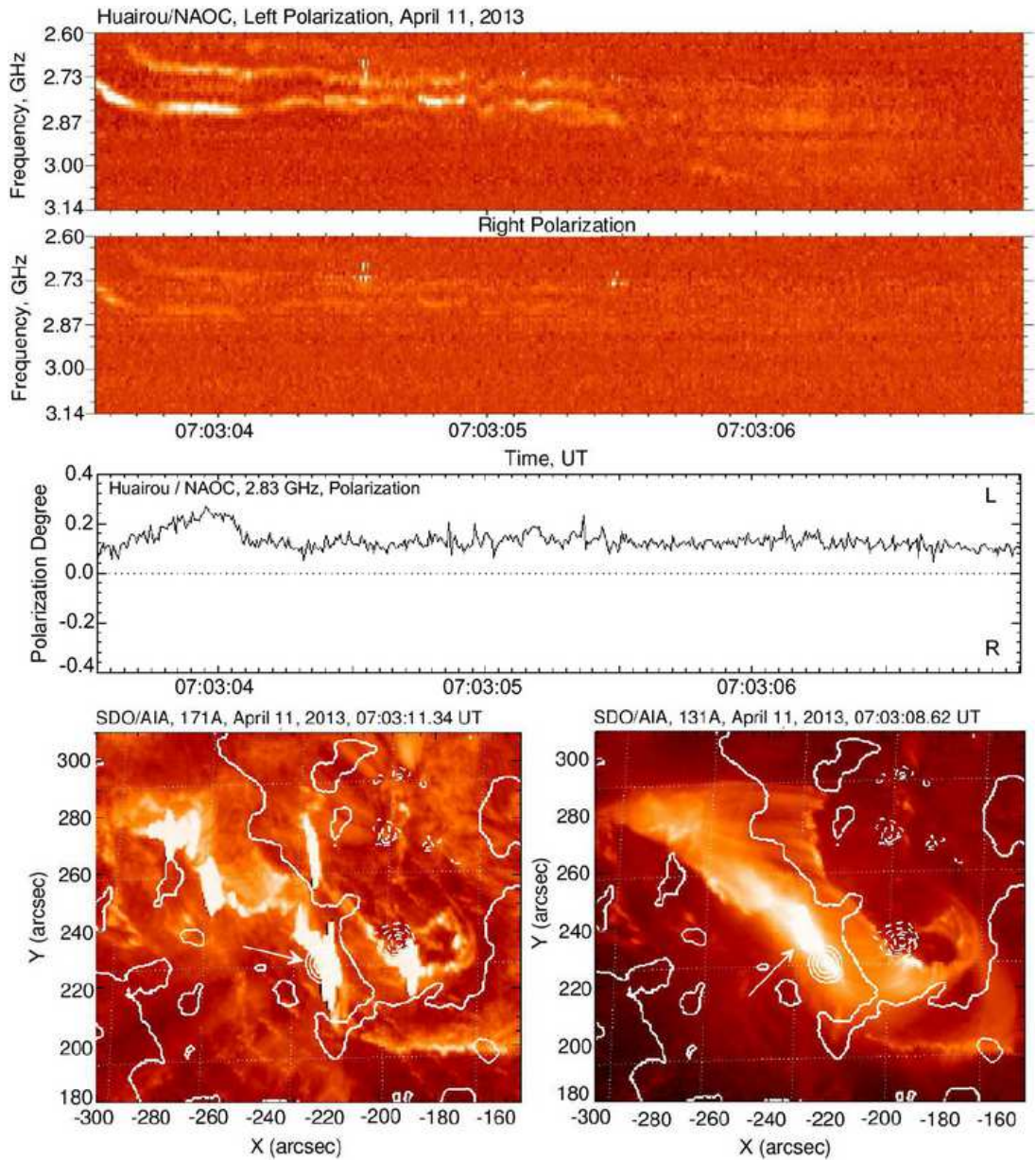}
 \caption{New ZP (the moment ZP5 in Fig. 3) in the Left polarization (about of $25 \pm5\%$)
 related to the new flare core shown by arrows in the left flare ribbon with N magnetic
 polarity (ordinary wave mode), according to two frames from the movie of SDO/AIA, 171 \AA~
(left) and 131 \AA~ (right).}
 \end{figure}

 \begin{figure}

 \centering
 \includegraphics[width=14cm, angle=0]{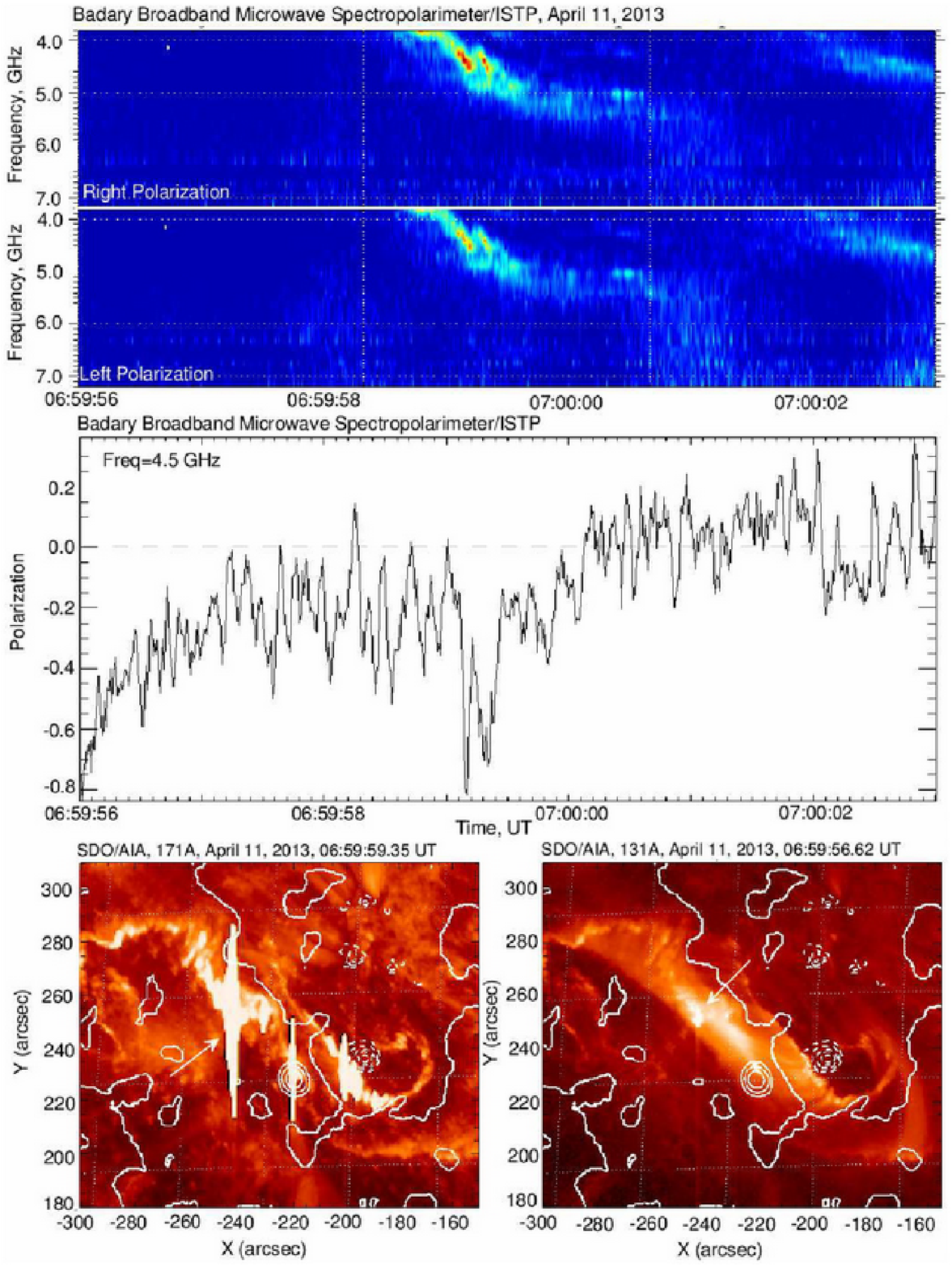}
 \caption{Fine structure observed with the BMS/Irkutsk in the range 4 – 7 GHz (top panel).
 Middle panel shows polarization profile at frequency 4.5 GHz. According to two frames in
 171 \AA~ and 131 \AA~ (two bottom panels) a new flare brightening takes place in the right
 flare ribbon with South magnetic polarity (ordinary wave). Two flare brightenings shown by
 the arrows in the bottom panels are responsible for the continuum emission with Left-handed
 polarization. }
 \end{figure}

 \section{DISCUSSION}

 The study of a short radio burst with rich fine structures on 13 April 2013 showed that each
 new radio maximum was related to a new flare brightening seen in EUV images of SDO/AIA. Each
 radio maximum has its own fine structures, usually composed of several stripes of ZP or ZP
 in the high frequency edge of fast pulsations. Such a relation indicates there is a close
 connection between radio sources of pulsations and ZP. The movie of SDO/AIA in 131 \AA~
 showed a flare loop arcade forming between two sigmoid flare ribbons. Therefore the flare
 dynamics consisted of consecutive magnetic reconnections in different arcade loops.

The polarization changed in accordance with the position of the new flare brightening.
The left flare ribbon was located above the North magnetic polarity (tail spot) and the
right ribbon above the South magnetic polarity (leading spot). In all cases, the radio
emission mode remained the ordinary one.  When the brightening took place at the looptops,
the polarization was very weak, almost zero.

  The magnetic field remained stable during the event. In the same time it was improbable
  to propose a motion of the radio source from one flare ribbon to another one during
  several seconds. Similar explanation of a gradual changing of the polarization sign
  at 17 GHz  (Nobeyama data) was proposed by Huang and Lin, 2006.

  Only a question arises: why we receive only a partial degree of polarization?~  If
  the emission is generated at the fundamental (by some mechanism) as the ordinary
  mode, it is fully polarized in the source. During propagation of the radio waves
  the observed polarization degree is changed due to a depolarization effect. The
  depolarization happens in a layer where the radio emission is propagated exactly
  across the magnetic field. In the considered event, the source geometry (the flare
  occurred at the disk centre) allows that this condition is possibly satisfied for
  a ZP source in a closed magnetic trap (for more details see Chernov, Zlobec (1995)).
The emission of fast radio pulsations is probably caused by fast electrons accelerated
in the upward direction in a vertical current sheet with the same period during magnetic
reconnection. Some of the fast particles that are accelerated downward can be captured
in a closed magnetic trap and they could be responsible for the emission of ZP by some
single mechanism. The diversity of ZP stripes is probably given by different conditions
in different arcade loops. This is natural scheme of flare processes which is in accordance
with the standard model of the solar flare as it is shown in Figure 10.

 \begin{figure}
 \centering
 \includegraphics[width=8cm, angle=0]{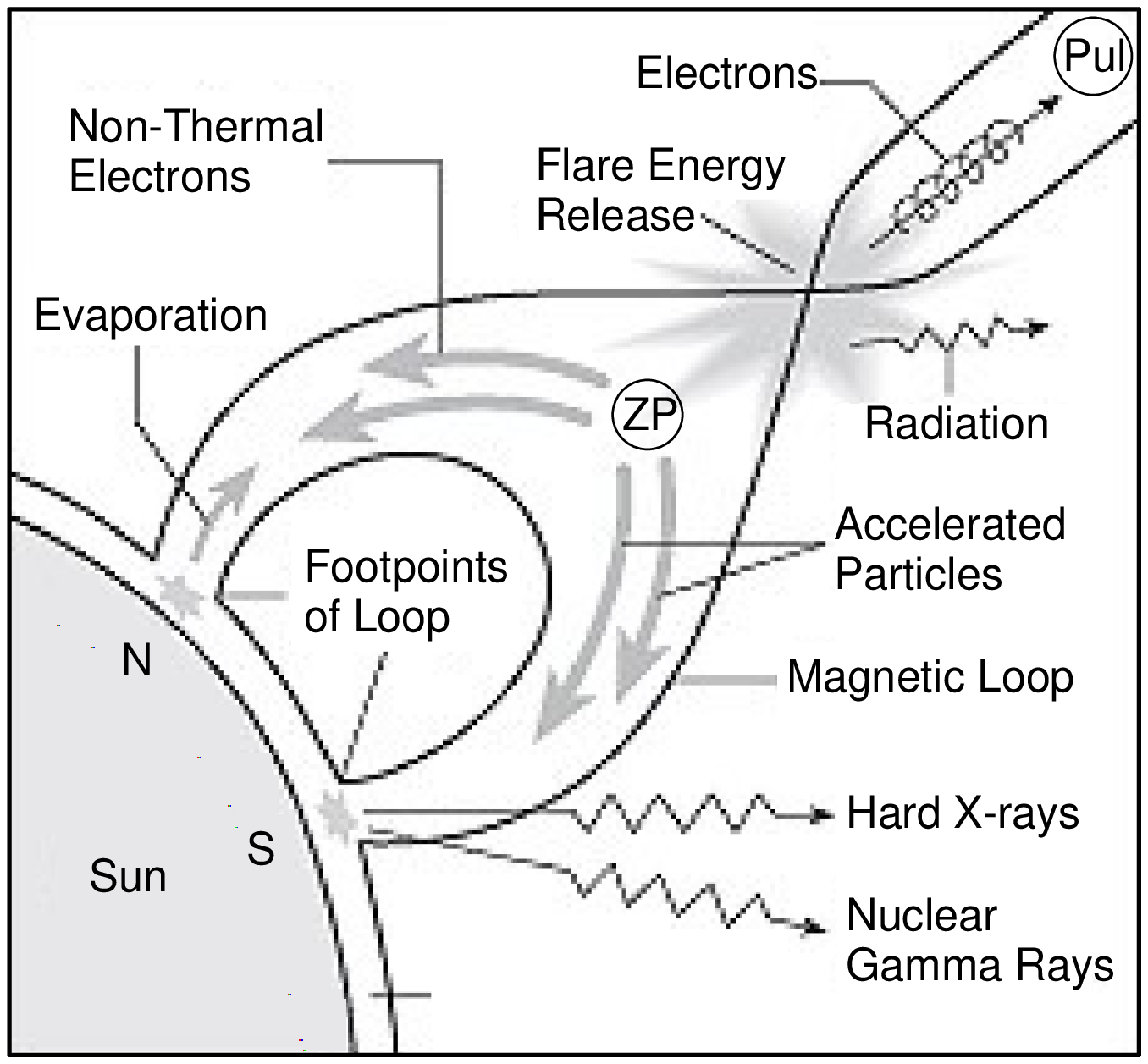}
 \caption{ Expected radio source positions of ZP and pulsations (Pul) in the scheme
 of the standard flare model of Aschwanden, 2006.}
 \end{figure}

      Such an expected position of a microwave ZP source at tops of flare magnetic loops
      was also confirmed in a new paper by Yasnov and Karlický (2015) in results of
      estimation of bremsstrahlung and cyclotron absorptions of radiation in the corona.

     We do not have any convincing evidence that the mechanism, generated ZP, is the
     emission of Bernstein modes as this was proposed by Tan et al. (2014) for the
     first strong ZP. First, an exact definition of the frequency separation between
     three (and sometimes four) zebra stripes is too problematic. Second, the polarization
     should be related to the extraordinary emission mode (Zlotnik, 1976; Kuznetsov, 2005).
     Third, we do not have any information about size of the radio source with height in
     the corona (distributed or point like).  Else, the radio emission defined by Bernstein
     modes must be weak, much weaker than in other mechanism, i.e., in the double plasma
     resonance (Zlotnik, 2009) or interaction of Langmuir waves with whistlers (Chernov, 2006).
     In both of the last models, the radio source should be distributed in heights, but in
     the double plasma resonance model a source should be stationary and in the whistler
     model – moving (depending on the group velocity of the whistler wave) (for more
     details see Chernov, Yan and Fu (2014). Furthemore, the spatial drift of ZP stripes
     should change synchronously with the changes of frequency drift in the dynamical spectrum.

      Recently, more than ten other mechanisms were proposed for ZP, but their significance
      remains uncertain (Chernov et al. 2014).
    To choose which mechanism applies, positional observations may be crucial, and it is
    desirable to observe a limb event. Now, we are expecting progress in the field of solar
    radio imaging spectroscopy. The first trial observations began on the new Chinese spectral
     radioheliograph (CSRH) which will be the largest and most advanced radio imaging
      telescope for the solar corona in the world (Yan et al. 2009; Yan et al. 2013).

 \section{Conclusions}
 \label{sect:Conclusion}

      We have shown that the polarization of ZP corresponds to the ordinary wave mode and
      it changes in accordance with dynamics of flare processes. Simultaneous or consecutive
       appearance of zebra-structure in different frequency ranges is obviously connected
       with the dynamics of flare processes.

     A future analysis needs to clarify whether a radio source showing ZP is really
     related to a closed magnetic loop, and if it is located at lower altitudes than
     the source of the pulsations, as expressed on the radio spectrum by ZP at the
     high frequency boundary of pulsations. New solar radio spectral imaging
     observations should help to compare the source sizes of different fine structures,
     and the main thing to determine whether the radio source does move.

     In the whistler model radio sources of fiber bursts and ZP are moving, and the
     spatial drift of ZP stripes should change synchronously with changes of the
     frequency drift in the dynamical spectrum. In the model of double plasma resonance,
     the ZP source must be rather stationary.

     It should be noted that the relative significance of several recent possible
     mechanisms remains uncertain.

 \begin{acknowledgements}

The authors are grateful to the RHESSI, SOHO (LASCO/EIT) and SDO teams for operating the
instruments and performing the basic data reduction, and especially, for the open data policy.
The research that was carried out by G.P. Chernov at National Astronomical Observatories (NAOC)
was supported by the Chinese Academy of Sciences Visiting Professorship for Senior International
 Scientists, grant No. 2011T1J20. The work of R. Sych was funded by Chinese Academy of Sciences
  President's International Fellowship Initiative, grant No. 2015VMA014. This work is supported
   by the Russian Foundation for Basic Research under Grants: 13-02-00044, 13-02-90472, 14-02-91157
   and 14-02-00367; NSFC grants: 11273030, 11103044, 11103039, 11221063, 11373039 and 113111042;
   MOST grant 2011CB811401; the National Major Scientific Equipment R\&D Project ZDYZ2009-3;
   and grant P209/12/00103 (GA CR).

 \end{acknowledgements}

\label{lastpage}

 \end{document}